\documentstyle[epsf,aps]{revtex}

\def \bi{\bibitem}

\def\nn{\nonumber}
\def\d{{\rm d}}
 \def\(({\left(}
 \def\)){\right)}
\def\bi{\bibitem}

\def \ov{\over}

\def \a{\alpha}
\def \b{\beta}

\def \d{{\rm d}}
\def \beq{\begin{equation}}
\def \eeq{\end{equation}}
\def \ln{{\rm ln}}
\def \ov{\over}

\def \a{\alpha}

\def \b{\beta}

\def \l{\left}
\def \ln{{\rm ln}}

\def \ab2{\alpha\beta^2}

\def\e{{\rm e}}
 \newcommand {\be} {\begin{equation}}
\newcommand {\bea} {\begin{eqnarray} \nonumber }
\newcommand {\ee} {\end{equation}}
\newcommand {\eea} {\end{eqnarray}}
 \newcommand {\eps} {\epsilon}

\newcommand {\si} {\sigma}

\newcommand {\Tr} {\mbox{Tr}}
\input{epsf}
\begin{document}
\title{Effective potential in glassy systems: theory and simulations}

\author{ Silvio Franz(*) and Giorgio Parisi(**)\\
}
\address{
(*) International Center for Theoretical Physics\\
Strada Costiera 11,
P.O. Box 563,
34100 Trieste (Italy)\\
and INFN Sezione di Trieste\\
(**) Universit\`a di Roma ``La Sapienza''\\
Piazzale A. Moro 2, 00185 Rome (Italy)\\
and INFN Sezione di Roma 1\\
e-mail: {\it franz@ictp.trieste.it, parisi@roma1.infn.it}
}
\date{November 1997}
\maketitle

\begin{abstract} 
We study the phase diagram of glassy systems in presence of an attractive coupling among
real replicas. We find competition among a localized and a delocalized phase, 
that are separated by a coexistence line as in ordinary first order phase transitions.  
The coexistence line terminates in a critical point.  
We present numerical simulations for binary glasses in which show that this scenario 
is realized.
\end{abstract}

 \vskip.5cm
\section{Introduction}

Supercooled liquids show a dramatic increase of
the relaxation time  as the 
temperature is lowered. The glassy transition 
is met at the temperature $T_g$ where
flowing becomes  not observable
  and the systems get out of equilibrium. 
One of the most suggestive ideas to 
rationalize this behavior can be found in the Gibbs-DiMarzio
theory and its elaborations \cite{vetro}.
This relates the observed finite time singularity 
to  a thermodynamic transition due to an entropy 
crisis, which would be  observable only in 
the infinite time limit.
Qualitatively the picture goes as follows. Still in the liquid phase, 
around a crossover temperature $T_c$,
the time scales for local motion of the molecules (vibrations) and large 
scale motion (diffusion) become widely separated, as if  
the system remained trapped for long times
in local minima of the free energy. The entropy $\Sigma$ 
associated to large scale motion,
often called configurational entropy or complexity,
can be related to the number  of free-energy minima ${\cal N}$
by the formula
${\cal N} =\exp(N\Sigma(T)$, ($N$ is the number of particles). 
As a function of temperature, $\Sigma$ is supposed to vanish 
at a finite temperature $T_s$, which  is the  ideal point 
of thermodynamic glass transition. The relaxation 
time is then related to the escape rate from the free-energy minima and 
one can argue in favor 
of Vogel-Fulcher like relations for the relaxation time as a function 
of the temperature with a divergence at $T_s$ \cite{kirtir,parisi}. 
As it was first realized by Kirkpatrick and 
Thirumalai \cite{kirtir},
the Gibbs-DiMarzio scenario is exactly implemented in a large 
class of infinite range disordered models, with the 
difference that the times needed to escape from  local
equilibrium states (and the corresponding 
free-energy barriers) diverge when the volume of the 
system goes to infinity. The value of $T_c$, which is to a large 
extent arbitrary in real systems, can be sharply defined in the 
mean-field limit. In fact this is the temperature where the
mode coupling theory \cite{MCT}, which is exact in these models, 
shows a divergent relaxation time. 
 On the contrary in 
short range systems $T_c$ signals a change in behavior but we cannot assign to it any sharply 
defined value.

It has been recently shown that if the models are generalized by introducing two coupled replicas of 
the same system one finds  that $T_c$ corresponds to the edge of a metastability
 region \cite{I}.  
In the same way 
the complexity is related to the difference of free-energy in the stable and in the metastable 
phase. In this paper we show how in the framework of coupled replicas the glass transition 
can be described as an ordinary phase transition. 
Enlarging the space of the parameters 
to include the coupling among replicas we find 
a first order transition line, terminating in 
a critical point.  Although our analysis is based 
on mean-field theory,  we will see that, as 
in ordinary first order phase transition, the  Maxwell construction will 
allow to extract the
qualitative features of the phase diagram of real systems.  
A sketch of these results 
has appeared in ref. \cite{letter}. To submit to test our picture in 
realistic systems, we have simulated coupled replicas of  binary mixtures with 
repulsive interaction \cite{HANSEN,pavetri}. 
These are known to vitrify for some values of the parameters defining the 
model. The results of Monte Carlo simulations     strongly support
the theoretical picture. 

This paper is structured as follows. 
In section II we present some general considerations on the effect of 
coupling replicas and we predict the behavior of a glass in the presence of two coupled replicas 
of the same system.  We have to distinguish two different case: the quenched and the annealed one, 
which have different properties. In section III we show that the previous arguments are indeed correct 
in a soluble model for the glassy transition, the $p$-spin spherical model. 
 In section IV we present 
the results of the  numerical simulations for the  binary mixtures. 
Finally in the last section we present our conclusions.

\section{Coupling replicas}
\subsection{The quenched case}

In this section we describe the construction of an effective potential 
in systems where long range order is absent, but that  
can remain stuck for a long time 
in metastable states. Generically, one can expect that
all the relevant metastable states at a given temperature
are equivalent as far as their macroscopic 
 characteristics are concerned.
Consequently, one can not identify any intrinsic order parameter allowing 
to distinguish one state 
from the others.  
In this situation it is possible to use 
as order parameter a degree of similarity among different 
points in configuration space. 
The procedure  is common in spin glass theory 
where the ``overlaps'' among 
different replicas appear as natural order parameters 
when one averages over the quenched disorder.
Here we discuss  an effective potential as 
a function of the overlap for generic systems,
which may or may not, as structural 
glasses, contain quenched disorder. 

Let us describe the construction in the 
case of a systems composed by only one type of particles 
with coordinates $x_i$, for $i=1,N$; the 
generalization to many kind of particles is trivial.  We consider two replicas of the same system, 
with coordinates $x$ and $y$ respectively, in an asymmetric relation.  The replica $y$ is a typical 
configuration distributed according to the Boltzmann-Gibbs law with the original Hamiltonian of the 
system    $H(y)$ at a temperature $T'$, and does not feel any influence from the replica $x$.
The replica $x$, instead, feels the influence of the replica $y$, and
for fixed value 
of $y$, thermalizes at a temperature $T$ with a
Hamiltonian
\be
H_\eps(x|y)=H(x)-\epsilon\sum_{i,k=1}^N w(x_i-y_k)
\ee

The function $w$ is different from zero only at short distance, an example is $w(x)=1$ if $|x|<a$ 
and $w(x)=0$ if $|x|>a$.  An interesting behavior is found when the value of $a$ is smaller that 
the typical interatomic distance (e.g.  $a=.3$ atomic distances).  The quantity $q(x,y)\equiv 
N^{-1}\sum_{i,k=1,N}w(x_i-y_k)$ measures then the similarity of the two configurations, and would be 
close to one when the two replicas stay in similar configurations.
Using the same terminology as in spin glasses \cite {EA,MPV,parisibook2} $q$ can be 
called the overlap of the two configurations. 
  For positive $\eps$ the $x$ 
variables feel then a potential which pushes them near to the $y$ variables.  We can define a 
free-energy for the $x$ variables in presence of the quenched configuration $y$  as:
\be
F(T,\eps,y)=(N\beta)^{-1}
\ln \left(\int dx
\exp{\{-\beta
H(x)+\beta\epsilon \sum_{i,k=1}^N w(x_i-y_k)\}}\right).
\label{def1}\ee
This quantity  should be self-averaging with respect to the distribution of the $y$ and can 
therefore be computed as
\be
F_Q(T,T',\epsilon)={\int dy \exp(-\beta' H(y)) F(T,\eps,y)
\over \int dy  \exp(-\beta' H(y))},
\label{def}
\ee
The temperature $T'$ of the reference configuration $y$ can be equal 
or different from that of the $x$ configuration ($T$). 

The free-energy (\ref{def}) is a well defined function that one can envisage to  evaluate
analytically or numerically. 
Its practical analytic evaluation can be performed with the aid of the 
replica method. For systems not containing quenched disorder one just needs 
to introduce replicas to average the logarithm in (\ref{def1}). This consists in 
 substituting $\log(Z)$ in (\ref{def1}) by $Z^m$ and evaluate $\langle \log(Z)\rangle=
\lim_{m\to 0}(\langle Z^m\rangle-1)/m$, where $Z$ is the argument  of the log in 
(\ref{def1}) and the angular brackets represent the average over the distribution of $y$. 
The formal 
procedure is similar to that used by Zippelius and coworkers to study vulcanization
\cite{zipp} and the one used by Given and Stell to for liquids in
 random quenched matrix \cite{given}. 
The physical meanings of ours and their constructions is however very different. 
Both in the vulcanization and in the liquid cases the replica method is used to 
average over some kind of  real quenched disorder,
the random crosslinking occurring at the
vulcanization transition in the first case and the  quenched matrix in the  
second. In our case, there is no quenched disorder. The coupling with the reference 
configuration $y$ is a theoretical tool that allows to probe regions 
of configuration space that have zero weight in the usual Boltzmann distribution, and 
that can  allow us to give a description of freezing even in absence 
of quenched disorder.

Writing explicitly the replicated partition function
\be
\langle
 Z^m \rangle=\int \d y \ \exp(-\b'H(y))\int \d x_1...\d x_n \ \exp\left[-\b
\sum_{a=1}^n H(x_a)+N\b\eps\sum_{a=1}^n q(y,x_a)\right]
\ee
we see that the problem is reduced to an $m+1$ component system in the limit 
$m\to 0$.  As we will see in the next section, the procedure needs to be
modified if the Hamiltonian $H$ contains some quenched parameters, to 
take into account the denominator in  
(\ref{def}).

Let us now try  understand qualitatively the behavior of $F_Q$ 
for small positive $\epsilon$ and  $T_0<T<T_c$, in the simpler 
case where the two temperatures are equal.  At $\epsilon=0$ the probability that the replica $x$ 
would stay in a same local minimum of the replica $y$ is exponentially small.  While, when 
$\epsilon>0$, the case in which the replica $x$ stay near to the replica $y$ is energetically 
favored.  The system can therefore stay in two different phases
\begin{itemize}
\item Replica $x$ different from $y$ ($q$ very small) and its  free-energy 
$F(T,\epsilon)\approx
F(T,0)$.
\item
 Replica $x$ near to $y$ (here $q\approx 1$).  The free-energy is given by $F(T,\epsilon)\approx 
 F(T,0) -\epsilon q +T\Sigma(T)$.
\end{itemize}

It is now clear that in this picture there is a first order phase transition at 
$\eps\approx T\Sigma(T)$ with a discontinuity in the internal energy given by $q$.  Moreover at 
$\epsilon=0$ the difference in free-energy among the two phases is exactly given by $T\Sigma(T)$.
For small $\eps$ one finds that the transition line starts as $T(\eps)=T_c+Const. \times \eps$.

The thermodynamic properties in the $T-\epsilon$ plane (for different values of $T'$) are quite 
interesting.  The previous argument tell us something only in the region of small $\epsilon$, the 
fate of the first order transition for large $\epsilon$ is a very interesting question.  In 
principle such a computation could be done in structural glasses by using the replicated hypernetted 
chain approach of \cite{MEPA} and work is in progress in this direction \cite{hnc}. 
As a first
investigation we limit ourselves to study what happens in a generalized spin glass model, the 
spherical $p$-spin models with long range forces \cite{pspin}.
We would like to stress another aspect that makes our approach interesting in connection with glass
physics.  Studying the usual Boltzmann measure of these models 
in the glassy phase one faces the 
problem that the configurations that give the dominant contribution at close 
but different temperatures look very different from each other (the so called 
chaotic temperature dependence of the measure). However, it happens in general  that 
metastable states that dominate the measure at a given temperature $T'$, that we call
$T'$-states, remain metastable if the 
temperature is changed \cite{bfp}. 
 With our method, if we fix $T'$ and 
we variate $T$ we can ``follow'' metastable states in temperature
In other words, 
we can modify the Boltzmann measure
so as to give non-vanishing weight to  the $T'$-states to different temperatures. 
This has important 
connections with cooling experiments in real systems.
At a given cooling rate, the system equilibrates within the (super-cooled)
liquid phase, until the glassy transition
temperature $T_g$ is reached. This is the last temperature where the system is able to equilibrate. 
Below that temperature genuine off-equilibrium phenomena as aging and memory effects set in the 
system. However,  one can expect that the
system remains confined for a long time 
in the metastable  state
 reached at $T_g$. Indeed this has been observed in recent numerical 
experiments in \cite{bako}. For short enough times after reached the temperature $T_g$ the 
system will be found
in local equilibrium in the ``analytic continuation'' of the state at $T_g$.
This hypothesis implies reversibility and can be valid only for  times
such that structural rearrangements can be neglected.
In this perspective one would like to define restricted Boltzmann-Gibbs 
measures in which only the configurations with a given distance from the quenched configuration 
$y$ have non-zero weight:
\be 
P(x|y)={1\ov Z(\b,y)} \e^{-\b H(x)}\delta(q(x,y)-q). 
\label{p}
\ee
As the constraint on the value of $q$  in (\ref{p}) is a global one, 
the free-energy associated to 
the distribution (\ref{p}), $V_Q(q)=-T \log Z(\b,y)$,  
is related to $F_Q(\eps)$ by  Legendre transform:
\be
V_Q(q) = \min_{\eps} F_Q(\eps)+\eps q.
\ee
This represents the minimal work required to keep the replica $x$ at fixed overlap $q$ 
from the replica $y$. In a situation with exponentially many minima, each one carrying 
vanishing contribution to the Boltzmann measure, one can expect a two minima structure 
of $V_Q$ at the mean field level. 
It is clear that a minimum should be found at the value of $q$ characterizing 
the typical overlap among different states. This corresponds to having the 
second replica in one of the exponentially many equilibrium states,  different from the
one where the first replica lie. In addition, if one does not allow for 
configurations
where $q$ is spatially inhomogeneous 
there is a minimum corresponding to the two replicas globally in the same 
state.\footnote{In real, finite dimensional
 systems the secondary minimum is washed out by the possibility of
having configurations with  inhomogeneous
$q$ but  the same global overlap  and smaller free energy then the 
homogeneous ones. This kind of configurations
with ``coexisting phases'' lead to conclude that, as in usual systems, the
potential has to be a convex function of $q$.} Therefore, for $T=T'$, 
the height of this secondary minimum 
with respect to the primary one has to be equal to the complexity $\Sigma(T)$ multiplied by $T$. 

\subsection{The annealed case}

A similar but different construction is the following: we consider two replicas
with total Hamiltonian
\be
H_\eps(x,y)=H(x)+H(y)-\epsilon\sum_{i,k=1,N}w(x_i-y_k)
\ee
The difference with the quenched case is that for $\eps\ne 0$ {\em both} $x$ and $y$ may not be  
equilibrium configurations. This construction, which we will see, gives similar results 
of the quenched one might be more advantageous in numerical simulations, where 
one can thermalize the two replicas at the same time.

Let see what we do expect in this case if we make the approximation that $q$ may be zero or $1$.
For $T>T_{c}$ there is a low $q$ phase in which the free energy and the internal energy are just 
the same as at $\eps=0$.
The most interesting phase is the high $q$ phase.
At $\eps=0$ the free energy in the region $T_{c}<T<T_{d}$ can be obtained for each of the two systems
by minimizing the total free energy
\be
F(f)=2[f-T\Sigma(f,T)]
\ee
If we couple the two systems and we restrict the analysis to  the pairs of configurations with 
$q\approx 1$, we have that
\be
F(\eps, f)=2f-T\Sigma(f,T)-\eps
\ee
In this case the minimum will be located at $\Sigma(f,T)=0$ also to temperatures higher than $T_{c}$,
i.e.  up to a temperature $T_{c,2}$ such that 
\be
T_{c,2}={\partial\Sigma(f,T)\over \partial T}{\Biggr |}_{T_{c,2}}
\ee

In other word the presence of term proportional to $\eps$ stabilizes the glass phase.  As far as the 
difference in free energy of the liquid phase and of the glassy phase is of order $(T-T_{c})^{2}$, 
we expect that the second order phase transition is transmuted into a first order one with at a 
temperature $T(\eps)=T_{c}+Const.\times
 \eps^{1/2}$ and a discontinuity in the internal energy proportional to 
$\eps^{1/2}$. This in contrast with the quenched case, where $T(\eps)$ is linear in $\eps$.
We see then, that although the symmetric coupling among replicas induces a 
kind of  deformation of 
the landscape, so that e.g. the various transition temperatures are changed, the global situation
is  similar to that of the quenched case with the main difference in the behavior of the transition 
line.

\section{An analytic computation}
\subsection{The $p$-spin spherical model}
The model that we are going to describe is the so called $p-$spin spherical model, which
has become in the last years one 
of the simple reference mean-field models for the structural glass transition \cite{pspin}.  
It will be clear from the 
form of the Hamiltonian
that this generalized spin glass is microscopically very different from a structural 
glass. The basis for its use as a model for the glass transition 
rely basically on the phenomenology of disordered models 
model with ``one step replica symmetry breaking transition'', which is strongly 
reminiscent to that of structural glasses. This issue, that has been discussed widely 
in the literature \cite{kirtir}, 
will be the starting point for the application of our discussion to real glasses. 
Just to mention a few facts, the model presents an ideal glassy transition to
a broken ergodicity phase with an extensive configurational  entropy, to a low temperature
zero complexity phase, very much like in the Gibbs-Di Marzio scenario, and its Langevin dynamics maps 
exactly in the schematic mode coupling theory \cite{vetro,MCT}
and its off-equilibrium generalization \cite{fh,pisa,parigi},
 capturing in this 
way many of the qualitative and some 
quantitative features of the supercooled and glassy relaxation. 
\begin{figure}
\begin{center}
\epsfxsize=350pt
\epsffile{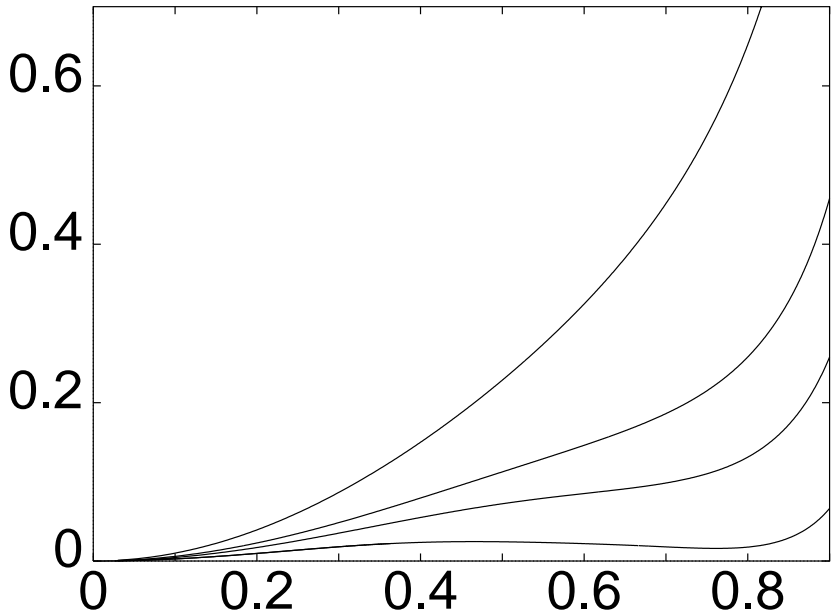} 
\end{center}
\caption[0]{\protect\label{figure1}  The
potential as a function of $q$ for
the p-spin model with $p=4$, $T'=0.523$ and various values of $T$, 
in order of decreasing temperatures from top to bottom.  
We have chosen $T_c<T'<T_s$. For $p=4$ one has $T_c=0.503$ and 
$T_s=0.544$. The curves have been normalized in such a way 
that $V_Q(0)=0$. } \end{figure}
Many study have shown explicitly that the random nature of the couplings in this models 
is not essential. In fact there have been found long range models with oscillatory couplings
and the same basic physics of the model we are going to describe \cite{mapari,boume,fh}.  
We than believe that many of the issues that we will discuss in this paper, and in particular 
the qualitatively features of the phase diagrams in the $T-\epsilon$ plane are quite universal and 
reflect very general properties of the phase space. We can thus conjecture that the phase diagram 
for real glasses is similar to that of the generalized spin glasses if we only consider the order of 
the phase transition and the topology of the various transition lines in the $T-\epsilon$ plane.

The model is defined in terms of $N$ real dynamical variables (spins) $S_i$, ($i=1,...,N$) subjected 
to the constraint $\sum_{i=1}^N S_i^2=N$ and interacting via the Hamiltonian 
\be
H_p=-\sum_{i_1<...<i_p}^{1,N} J_{i_1,...,i_p}S_{i_1}...S_{i_p}
\label{Hp}
\ee
with independent centered Gaussian 
couplings $J_{i_1,...,i_p}$ with variance $\overline{J_{i_1,...,i_p}^2}=p!/(2 N^{p-1})$.  
The model for $p>3$ has a one step replica braking, the
transition pattern which has been repedetly
shown to be deeply related to the Gibbs-Di Marzio entropy crisis mechanism \cite{kirtir,remi}.

In spin models the natural way to couple two replicas consist in adding to the Hamiltonian a term 
$-\eps\sum_i S_i {S'_i}$ and with the usual definition 
of the overlap $q=N^{-1}\sum_{i=1,N}S_i {S'_i}$.  The overlap $q$ is 
equal to one if the configurations of the two systems coincide.   

\subsection{The Quenched case}
The quenched potential for the model is obtained inserting the Hamiltonian (\ref{Hp})
in the  the general definition 
(\ref{def}). 
The two replicas potential is
\bea
F_Q(T,T',\epsilon)&=&\overline{{\int \d S' \exp(-\beta' H(S')) F(T,\eps,S')
\over \int \d S'  \exp(-\beta' H(S'))}},\label{fq}\\
F(T,\eps,y)&=&(N\beta)^{-1}
 \ln \left(\int dS \exp\{-\beta
H(S)+\beta\epsilon \sum_{i,k=1,N}S_k S'_k \}\right).\nonumber
\eea
where the bar denotes the average over the $J$'s.
In addition to the $m$ replicas needed to average the logarithm $\overline{\log(Z)}
=\lim_{m\to 0} (\overline{Z^m}-1)/n$, 
in order to average 
over the quenched parameters $J_{i_1,...,i_p}$ we also need to represent 
the denominator $1/z$ in (\ref{fq}) as $\lim_{n->0}z^{n-1}$.
As usual the computation is performed continuing from integer $n$ and $m$. 
We have then $n$ unconstrained replicas $S_i^a$, ($a=1,...,n$) and $m$ constrained replicas 
$S_i^\a$, ($\a=1,...,m$).
The average over the quenched disorder induces a coupling among replicas, and as usual 
in this models, the
order parameter for the theory is the matrix of the overlaps among all the $n+m$ replicas
of the system. This can be conveniently arranged in three matrices describing respectively 
the overlaps of the unconstrained replicas, $Q_{a,b}=1/N \sum_i S_i^a S_i^b$, 
($a,b=1,...,n$) the overlap
among the constrained replicas $R_{\a,\b}= 1/N \sum_i S_i^\a S_i^\b$, ($\a,\b=1,...,m$) and the mutual 
overlap among constrained and unconstrained replicas $P_{a,\a}=1/N \sum_i S_i^a S_i^\a$. 
In terms of these order parameters one finds
\bea
{1\ov N}\log Z_2^{(n,m)}= & &
{1\ov 2}\l[\sum_{a,b}^{1,n} \b'^2 f(Q_{a,b})+\sum_{\a,\b}^{1,m} \b^2
f(R_{\a,\b})+2\sum_{a,\a} \b\b' f(P_{a,\a})\right] \nn\\
& &
+2\b\eps\sum_{\a=1}^m P_{1\a}
+{1\ov 2}\Tr \log \left(\begin{array}{cc}
Q & P \\
P^T & R
\end{array}
\right)
\label{25}
\eea
where we have written $f(q)=q^p/2$.
From (\ref{25}) one has to extract the terms proportional to $m$ in order to evaluate the 
effective potential. 
The procedure that we have sketched here has been  
discussed in  ref. \cite{I,AIG,bfp}, to which we address the 
interested reader for the details. 
In the following we will concentrate to  temperatures $T>T_s$. In this region 
the matrix describing the unconstrained replicas has the simple form 
$Q_{a,b}=\delta_{a,b}$, which is valid both in the paramagnetic phase for $T>T_c$ 
and in the non ergodic phase for 
$T_s<T<T_c$ with exponentially many states with 
vanishing weights. 
\begin{figure} 
\begin{center}
\epsfxsize=350pt
\end{center}
\epsffile{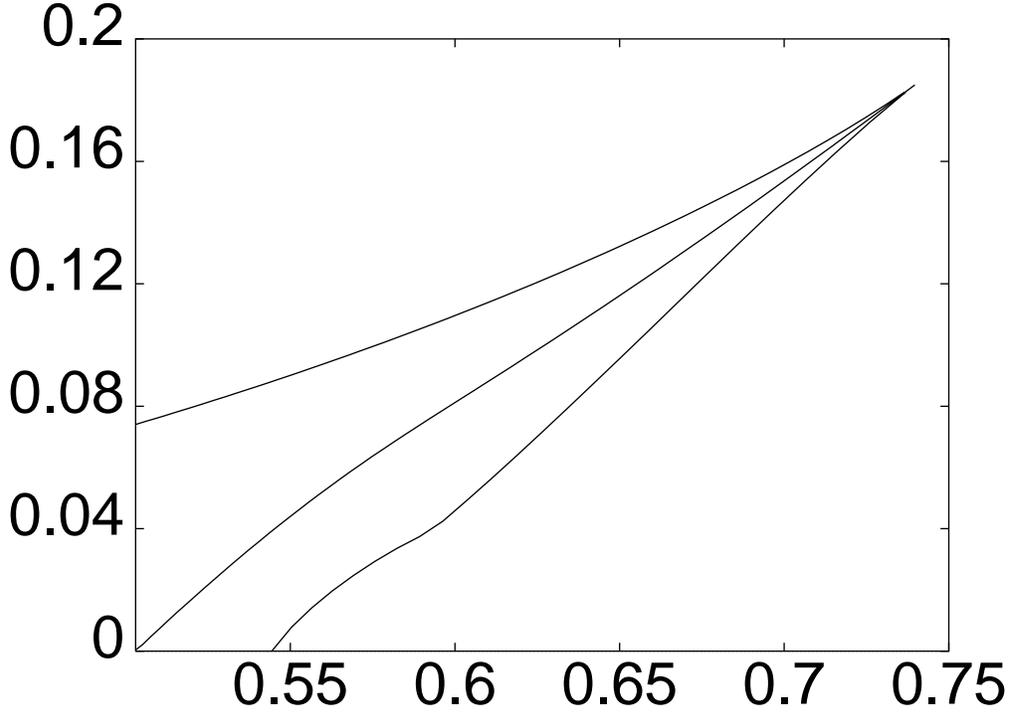}   
\caption[0]{\protect\label{figure2} Phase diagram in the $\eps-T$ 
plane for $p=4$ and  $T'=T$.  The upper curve is the spinodal of the low $q$
phase, the  lower one the spinodal of the high $q$ 
state, and the middle curve the coexistence line. The coexistence 
line touches the axes $\eps=0$ at $T=T_s$, while the spinodal of 
the high $q$ phase touches it at $T=T_c$. For $T<T_s$ the 
spinodal of the low $q$ phase  remains finite and touches the 
$T=0$ axes at finite $\eps$.} 
\end{figure} 
In this conditions the constrained replicas have non 
zero overlap only with replica $a=1$ and therefore, 
$P_{a,\a}=\delta_{a,1} q$. For the matrix $R$ the most general ansatz needed 
here is 
a ``one step broken'' structure \cite{MPV}. Both the structure and the 
physical meaning of this ansatz have  been discussed widely in the literature, and
in standard notations we  parameterize it  by the  three parameters $q_0,q_1,x$.

In the following we will study the phase diagram of the model in the $\eps -T$ plane in two 
situations: a) $T'=T$, corresponding to restricting the partition sum to the vicinity of a 
particular equilibrium state at each temperature.  b) $T'$ fixed, corresponding to probe the 
evolution of the free-energy landscape in the vicinity of a fixed configuration of equilibrium at 
$T'$ when $T$ is changed. 
The Legendre transform of $F_Q(T,T',\eps)$, $V_Q(q,T,T')\equiv\min_\eps 
F_Q(T,T',\eps)+\eps q$, admits the following expression 
in terms of the variational parameters defined above. 
\bea
V_Q (q) = -\frac{1}{2\beta}  \left\{
2\beta \beta' f(q) - \beta^2 \(( (1-x) f(q_1) + x f(q_0) \))
+\frac{x-1}{x}\ln(1-q_1) \right. \\
\left. + \frac{1}{x} 
\ln \((1-(1-x) q_1 -x q_0 \))
+\frac{q_0 - q^2}{1-(1-x) q_1 -x q_0} \right\}
\eea
where $V_Q$ has to be maximized with respect to $q_0$, $q_1$ and
$x$. Depending on the values of $\b,\b'$ and $q$, the solution 
of the saddle point equations can be either replica symmetric 
with $x=0$ or $x=1$, or display replica symmetry breaking 
with $q_1\ne q_0$, and $x\ne 0,1$  \cite{bfp}. 

We see from fig. \ref{figure1} that the shape of the function $V$ is 
the characteristic one of a mean-field system 
undergoing a first order phase transition.  At high enough temperature $V_Q$ is an increasing and 
convex function of $q$ with a single minimum for $q=0$.  Decreasing the temperature, we find
 a value 
$T_{cr}$, where for the first time a point $q_{cr}$ with $V_Q''(q_{cr})=0$ appears. 
The potential looses the 
convexity property and for $T\leq T_{cr}$
a phase transition can be induced by a field.  A secondary minimum develops 
at $T_c$, the temperature of dynamical transition \cite{kirtir}, signaling the presence of long-life 
metastable states.  The minimum of the potential has received a dynamical interpretation
in  
\cite{I,BBM,bfp} where it has been shown that its characteristics (internal energy, 
self-overlap, etc.)
correspond to   the states reached at long times by the evolution at temperature 
$T$ starting at an initial time from an equilibrium configuration at temperature $T'$.
 In figure \ref{figure1} we show 
the shape of the potential in the various regions.  
\begin{figure} 
\begin{center}
\epsfxsize=350pt
\end{center}
\epsffile{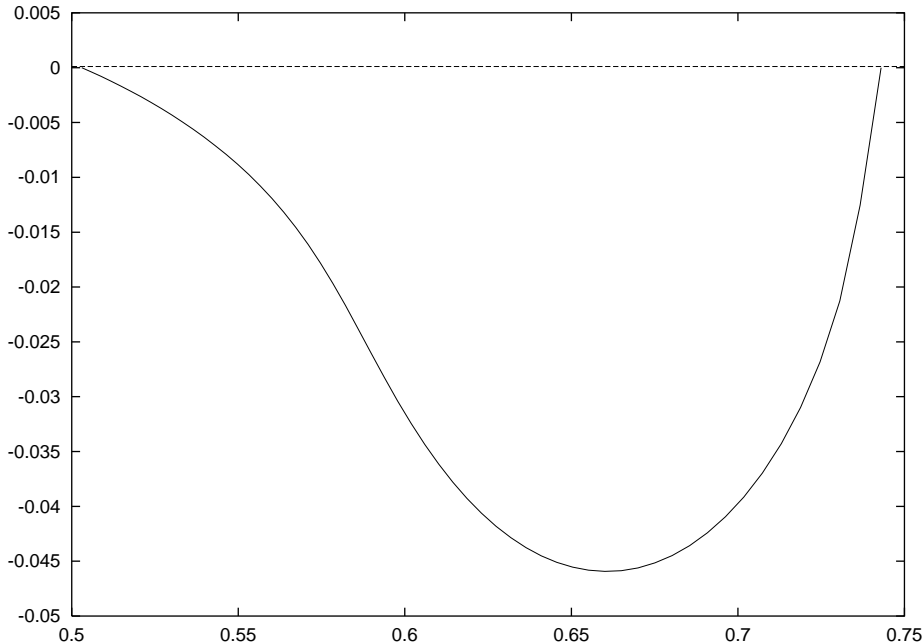}   
\caption[0]{\protect\label{cvtt} Latent heat of the transition as a 
function of the temperature  for $p=4$ and $T'=T$.   } 
\end{figure} 
Although the behavior of the potential function is analogous to the one found in ordinary systems 
undergoing a first order phase transition the interpretation is here radically different.  While in 
ordinary cases different minima represent qualitatively different thermodynamical states (e.g.  gas 
and liquid), this is not the case in the potential discussed here.  In our problem the local minimum 
appears when ergodicity  breaks,
 and the configuration space splits into an exponentially large 
number of components.  The two minima are different manifestations of the existence 
of exponentially many states with similar 
characteristics. In the high $q$ minimum the system $S$ is in the same microscopic state as $S'$,
while in the low $q$ minimum it can be in any other state among the exponentially many 
that contribute to the Boltzmann measure.  
The height of the secondary minimum, relative to the one at $q=0$ measures the 
free-energy loss to keep the system in the same component of the quenched one.  At equal 
temperatures $T=T'$ this is just equal to 
the complexity $\Sigma$ multiplied by $T$ and it 
 goes to zero at $T_s$, where coexistence in 
zero coupling takes place without a release of latent heat and both minima lie 
at the same height. 
For $T\neq T'$ the height of the secondary minimum  also takes into account the free-energy 
variation of the equilibrium state at temperature $T'$ when ``followed'' (i.e. 
adiabatically cooled or heated) at temperature $T$.  

The 
presence of the field $\eps$ adds finite stability to the metastable states, 
and the transition is 
displaced to higher temperatures. The position of the transition line can be 
computed via the Maxwell
construction. 
In figure \ref{figure2}
 we display the phase diagram of the $p=4$ model in 
the case $T'=T$.  The coexistence line departs from the axes $\eps=0$ at the transition temperature 
$T_s$ and reaches monotonically a critical point $(T_{cr},\eps_{cr})$. For the mean field 
model under study one can see that the exponents characterizing 
the critical point  are the classical ones. 
  In figure \ref{figure2} we  also show 
the spinodal of the high $q$ solution, which touches the $\eps=0$ axes at the dynamical temperature 
$T_c$, and the spinodal of the low $q$ solution for temperatures larger then $T_s$.  
While the transition in zero field is not 
accompanied by heat release, a latent heat is present in non zero $\eps$.  In figure \ref{cvtt}
 we show, in 
the same conditions of fig. \ref{figure2},
 the latent heat ${\cal Q}=E_+-E_- -\eps(q_+-q_-)$ where $E_+$ 
($q_+$) and $E_-$ ($q_- $) are the internal energies (overlaps) respectively of the high and low $q$ 
phases. Notice that  the latent heat vanishes  at the critical 
point (as it should), and at $T=0$.
The coexistence 
line for $T'$ fixed, in the interval $T_c\leq T'\leq T_s$ is qualitatively similar to the one of 
figure \ref{figure2}
 at high enough temperature, but (for $T'> T_s$) it never touches the axes $\eps=0$.
\begin{figure}
\begin{center}
\epsfxsize=350pt
\end{center}
 \epsffile{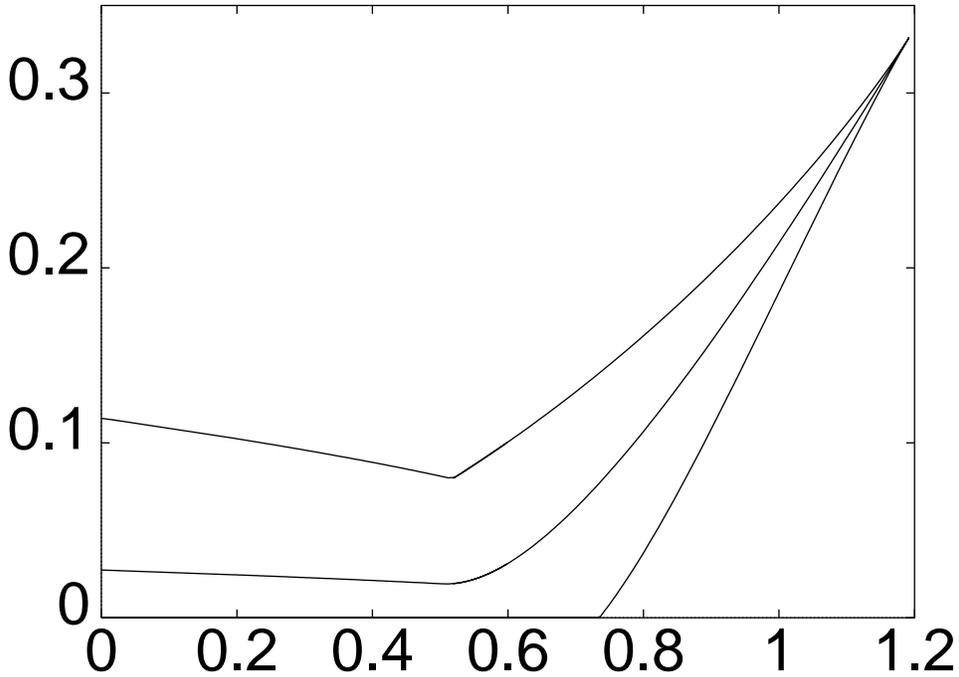}   
\caption[0]{\protect\label{figure3} Phase diagram in the $\eps-T$ 
plane for $p=4$ and $T'=0.523$.  The upper curve is the 
spinodal of the low $q$ phase, the  lower one the spinodal of the high $q$ 
phase, and the middle curve the coexistence line.    } 
\end{figure}  
Even at zero temperature there is a first order phase transition in $\eps$, reflecting the fact that 
the ground state of the system is lower then the energy of the reference state when followed at 
$T=0$.  This can be seen in figure \ref{figure3}
 where we show the phase diagram for a value of $T'$ such that 
  $T_c<T'<T_s$.  
In figure \ref{figure4} we see that the latent heat for $T'\ne T$ is qualitatively similar to 
the one for $T'=T$ at high temperature, while at low temperature it change sign
and becomes zero only at $T=0$. 
\begin{figure} 
\begin{center}
\epsfxsize=350pt
\end{center}
\epsffile{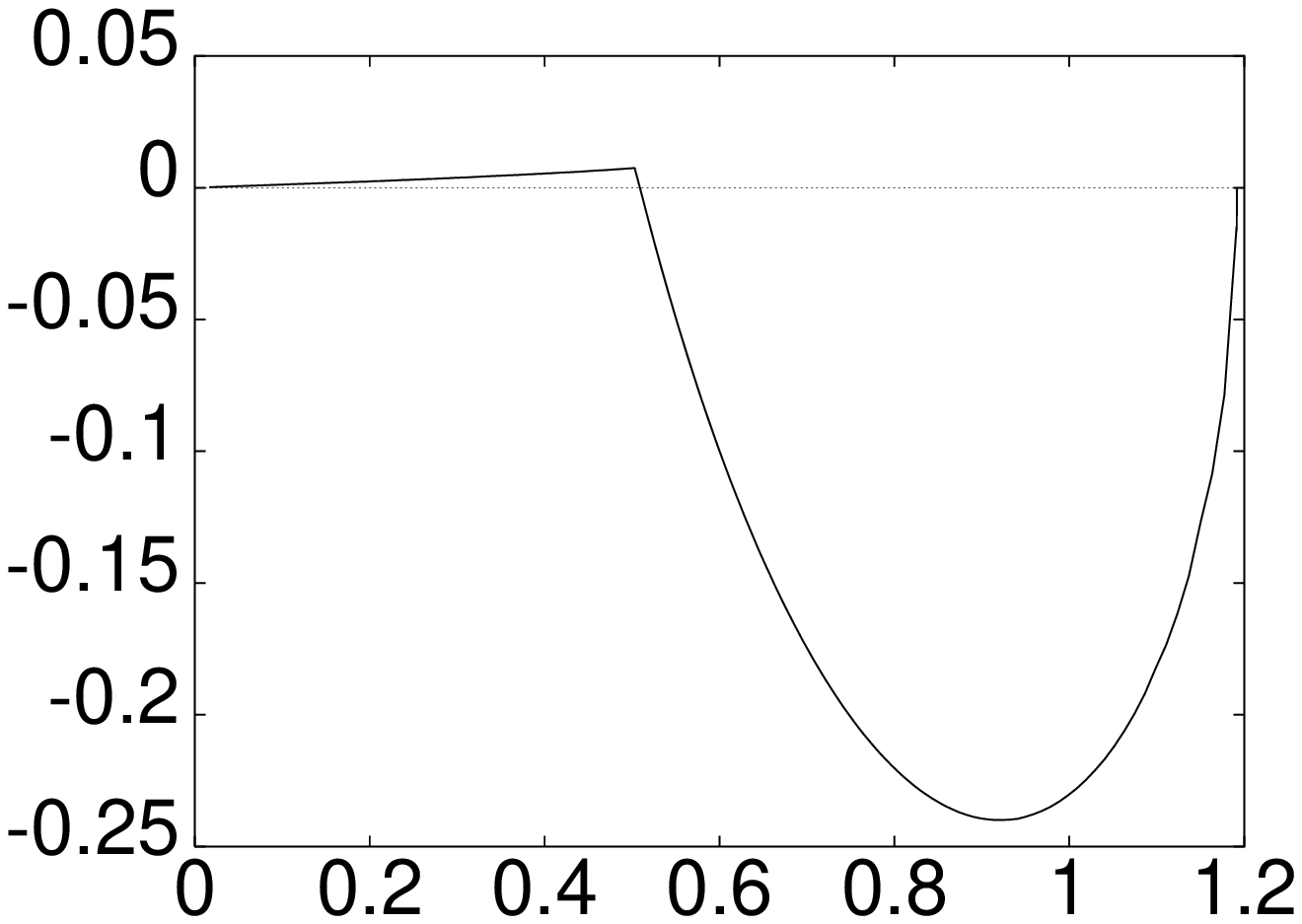}   
\caption[0]{\protect\label{figure4} Latent heat of the transition as a 
function of the temperature  for $p$ and $T'$ as in fig. \ref{figure3}.   } 
\end{figure} 
  The high $q$ 
phase roughly reflects the properties of the equilibrium states at temperature $T'$ followed at 
temperature $T$, while the low $q$ phase reflects the properties of the true equilibrium states at 
temperature $T$.  At high temperature the high $q$ phase is energetically favored, while 
at low temperature it has an energy higher then the one of equilibrium.  
The point where ${\cal Q}$ 
changes sign reflects this fact, and 
does not correspond to a second order phase transition.  Finally, in figure \ref{figure5}
 we show, 
for a fixed temperature the curve of $q(\eps)$ obtained by the Maxwell construction.

In closing this section we would like to comment on the use of the Maxwell 
construction for finite dimensional systems. In ordinary systems, the 
justification of  the Maxwell construction is in the phenomenon of
 phase coexistence. 
Here we do not know what a supposed coexistence in physical space 
of the
high $q$ and the low $q$ 
phase would mean. This, together with the related problem of 
of finding solutions with inhomogeneous $q$ 
 to the mean-field equations in finite dimensional models is an 
open problem and we 
let it for further investigation. For the time being we limit 
ourselves to look for  support for our construction 
in numerical simulations. This will be the aim of next section. 
\begin{figure} 
\begin{center}
\epsfxsize=350pt
\epsffile{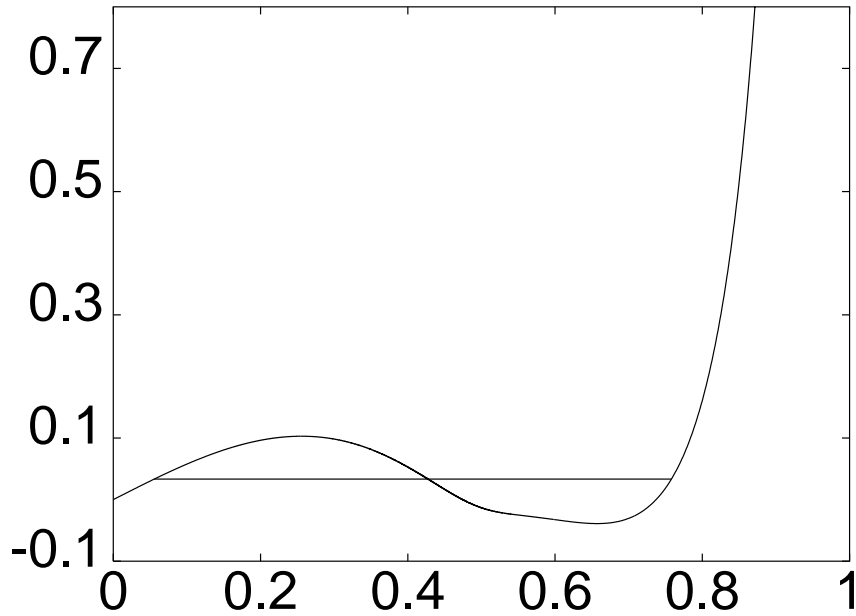} 
\end{center}
 \caption[0]{\protect\label{figure5} Equation 
of state for $p=4$, $T'=0.523$ and $T=0.609$.  The 
horizontal line corresponds to coexistence and is obtained by the 
Maxwell construction.} 
\end{figure} 
\subsection{The Annealed case}
\begin{figure} \epsfxsize=350pt
\epsffile{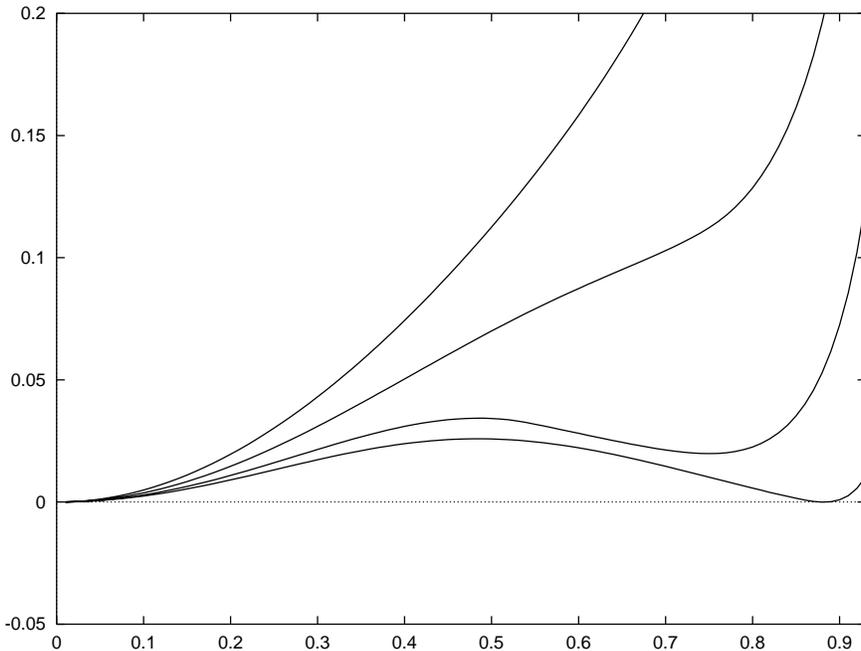}    \caption[0]{\protect\label{ann_pot}  The
annealed potential $V_A(q)$ as a function of $q$ for $p=4$
and various values of $T$, 
in order of decreasing temperatures from top to bottom.  For $p=4$ one has $T_c=0.503$ and 
$T_s=0.544$. The curves are normalized to have $V_A(0)=0$. } \end{figure}

The aim of this section is to show rapidly that the annealed potential gives results
qualitatively  similar to the quenched 
one. The detailed analysis can be found in \cite{KPV}. 
 In this case we want to compute the average of the free energy of the two coupled replicas
 over the 
quenched couplings. 
The procedure to do this with replicas is explained in \cite{FPVI,KPV}. In this cases 
there are the overlaps describing the system can be conveniently 
arranged into 3 replica matrices $Q_{a,b}=1/N\sum_i S_i^a S_i^b$,
$Q'_{a,b}=1/N\sum_i {S'_i}^a {S'_i}^b$ and $P_{a,b}=1/N\sum_i S_i^a {S'_i}^b$.
All the matrices have the dimension
$n\times n$ and the symmetry among the two system implies $Q=Q'$ and $P_{a,b}=P_{b,a}$. 
The most general ansatz needed in the problem is a ``one step replica symmetry breaking'' form
where the matrices $Q$ and $P$ are parameterized respectively by the parameters 
$(q_1,q_0=0,x)$ and $(\tilde{p},p_1,p_0=0,x)$. The resulting free-energy is 
\bea
V_A(q)&=&- \beta\,\left[ f(1) + f(q) - 
       \left( 1 - x \right) \,\left( f(p_1) + f(q_1) \right) 
        \right] \\ & & - 
  {\frac 1 {2\,\beta}}{\left( 1 - {\frac{1}{x}} \right) \,
      \left[ \log (1 + p_1 -q - q_1) + 
        \log (1 - p_1 + q - q_1) \right] }\nn\\ & & - 
  {\frac 1 {2\,\beta\,x}}\left[{\log (1 + p_1 - q - q_1 + 
        \left( -p_1 + q_1 \right) \,x) +}\right.\nn \\ 
   & &   \left.{\log (1 - p_1 + q - q_1 + 
        \left( p_1 + q_1 \right) \,x)}\right]
\eea
which has to be optimized with respect to $q_1,p_1$ and $x$. 
In figure \ref{ann_pot} we see that qualitatively the situation resembles to the quenched case, with
a non convex potential  function at low enough temperature. It is clear that
the essential features of the phase diagram or the previous section with
a first order transition line terminating in a critical point are present also in 
this case, although the actual values of the different characteristic temperatures 
will be different. 

\section{Real glasses}
\subsection{The Model}
We have  tested  the ideas presented in this note for binary fluids.  
Although we did not perform a systematic study of the phase transition line, we found evidence 
for a first order transition in presence of coupling among replicas both in the annealed and
in the quenched case. 
The model we consider is 
the following.  We  take  a mixture of soft particle of different sizes.  Half of the particles 
are of type $A$, half of type $B$ and the interaction among the particle is given by
\begin{equation}
H(x)= \sum_{{i<k}} \left(\frac{\si(i)+\si(k)}{|x_{i}-x_{k}|}\right)^{12},
\label{HAMILTONIAN}
\end{equation}
where the radii ($\si(i)$) depend on the type of particles.  This model has been 
carefully studied studied in the past \cite{HANSEN}.  It is 
known that a choice of the radii such that $\si_{B}/\si_{A}=1.2$ strongly 
inhibits crystallization and the systems goes into a glassy phase when it is 
cooled.  Using the same conventions of the previous investigators we consider 
particles of average diameter $1$. More precisely we set
\begin{equation} 
{\si_{A}^{3}+ 2 (\si_{A}+\si_{B})^{3}+\si_{B}^{3}\over 4}=1.
\label{RAGGI}
\end{equation}
 Due to the simple scaling behavior of the potential, the thermodynamic quantities depend 
only on the quantity $T^{4}/ \rho$, $T$ and $\rho$ being respectively the temperature and the 
density.  For definiteness we have taken $\rho=1$.

This is one of the simplest models of glass forming materials and we have chosen it because of 
its simplicity. The model as been widely studied especially for this choice of the parameters.  It is 
usual to introduce the quantity $\Gamma
\equiv
\beta^{4}$.  The glass transition is known to happen around $\Gamma=1.45$ \cite{HANSEN}.  It has 
been shown that aging appears below this temperature, and that the the time dependent correlation 
and response functions are well in agreement with the prediction of one step replica symmetry 
breaking below this temperature \cite{pavetri}.

Our simulation are done using a Monte Carlo algorithm, which is easier to 
deal with than molecular dynamics.
  Each particle is shifted by a random amount at each step, and the size of 
the shift is fixed by the condition that the average acceptance rate of the 
proposal change is about .5.  Particles are placed in a cubic box with periodic 
boundary conditions.

Following the discussion in the introduction, we have introduced two copies $x$ and $y$ of the same 
system e have introduced the quantity $q$ defined as
\be q(x,y)\equiv{1\over N} \sum_{i,k }^{1,N} w(x_{i}-y_{k}),
\ee
where the sum over $i$ and $k$ runs over all possible $N^{2}/2$ 
pairs of particles of the same kind.
 The function $w$ is chosen  in such a way that the quantity $q$ counts the percentage 
of particles such that there is a similar particle nearby in the other configuration.  
The form we 
have considered is
\be w(x)={a^{12} \over x^{12}+a^{12}}, \ee 
with $a=.3$ The function $w$ is very small when $x>>.3$ and near to $1$ for $x<.3$.  The value of 
$q$ will thus be a number very near to $1$ for similar configurations (in which the particles have 
moved of less than $a$) and it will be much smaller value (less than .1) for unrelated 
configurations.  The value of $a$ has been chosen in such a way that 
$q$ is insensitive to thermal fluctuations.

\subsection{Numerical results for the quenched case}
In this case the replica $y$ is at equilibrium with the Hamiltonian $H(y)$ while the Hamiltonian 
of the replica $x$ is
\be
H(x|y)=H(x)-N\tilde\eps q(x,y)
\label{Hxy}
\ee
where $\beta \tilde\eps = \eps$.

Our aim would be to find out if there is a first order transition in the plane $\eps - T$ and to 
locate the transition line.  In principle it is rather difficult to find out the precise position of 
a first order phase transition. 
 The reason is quite simple: in a dynamical simulation the mean life in a metastable phase is 
exponentially large just near the transition.

We have first followed an exploratory approach by monitoring the 
properties of the system as function 
of the coupling, temperature and their time variation, 
 when we go from one phase to the other.  In this case the results will be 
function of the speed at which we change the parameters
 and the precise value of the phase transition 
point will be obtained only in the limit of zero speed.

This gives an approximate information on the position of the transition. 
We have taken a system with 66 particles  and we have thermalized it at a given 
value of $\Gamma$ for $2^k$ Monte Carlo sweeps
(we have data for $k=7-17$ in order to estimate the $k$ dependence). 
\begin{figure}
\begin{center}
\epsfxsize=350pt
\epsffile{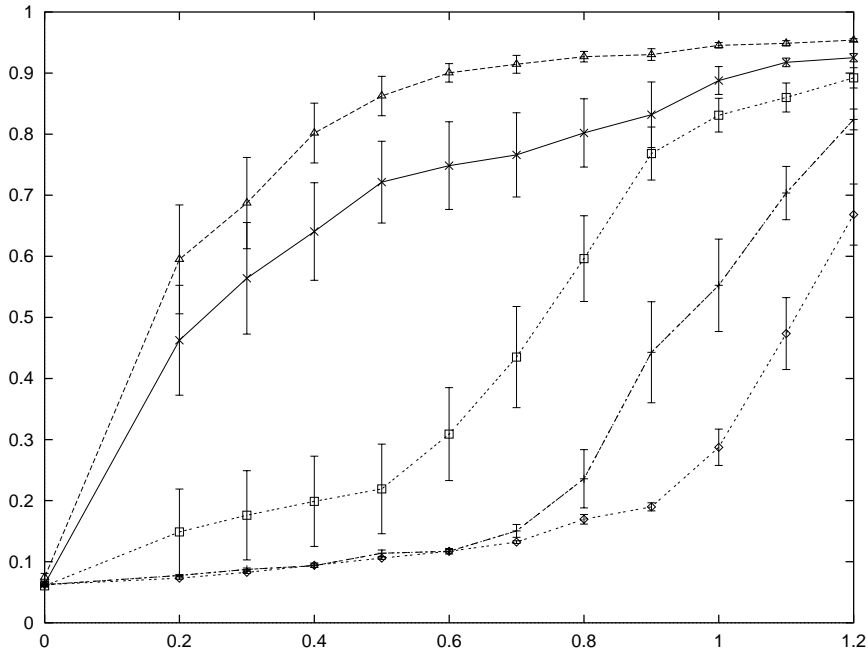} 
\end{center}
   \caption[0]{\protect\label{qeps}  
The overlap as a function of the coupling for various values of $\Gamma$, 
from  bottom  to top $\Gamma=1.25,1.30,1.35,1.40,1.45$. The number of particles
is 66, the value of $q$ is averaged over 10 different simulations, and the
number of Monte Carlo step for each value of $\eps$ in each simulation is
$2^{17}$. }
 \end{figure}
 At this point we start
the evolution of a second system which is coupled to the configuration reached by the first one ($y$) 
as in (\ref{Hxy}) with a value of $\eps=1.2$, taking as  initial condition 
the quenched configuration $y$. We let then the system thermalize for other $2^{k}$
steps and we measure the overlap in the last quarter of the run. Starting from the final configuration we 
decrease the value of $\eps$ by $0.1$ and we perform the  $2^k$ Monte Carlo sweeps. This procedure is
repeated up to $\eps=0$.

\begin{figure}
\begin{center}
\epsfxsize=350pt
\epsffile{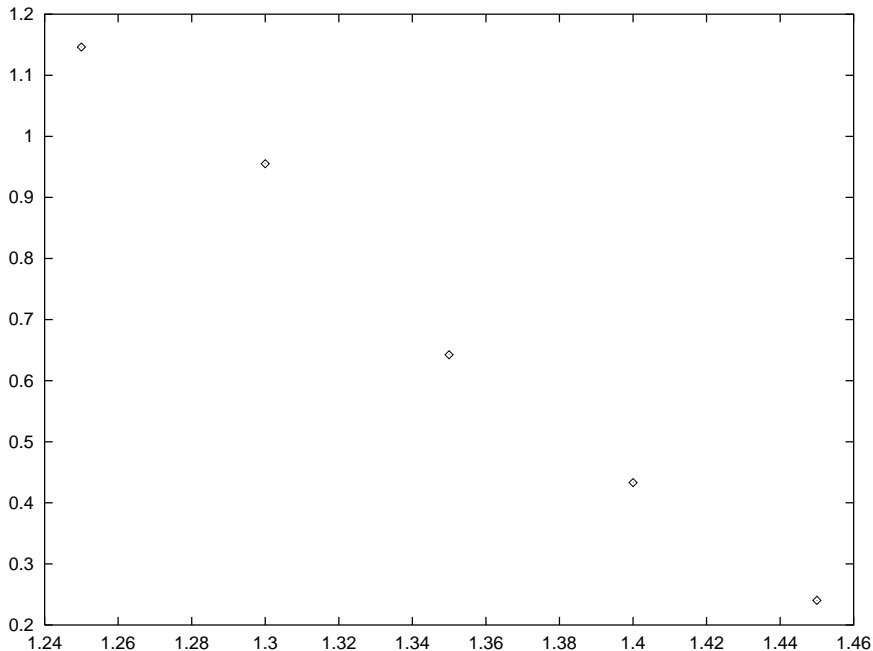} 
\end{center}
   \caption[0]{\protect\label{pha}  
Rough estimate of the transition line in the $\eps -\Gamma$ plane, obtained 
as the value of $\eps$ at which the overlap shown in the previous figure is 
equal to 0.7.}
 \end{figure}

In figure \ref{qeps} we present the data relative of this procedure for $k=17$. We see that 
at high temperature (low $\Gamma$) the value of $q$ drops to nearly  zero at already 
high values of $\eps$, while for lower temperature, it persists to high values 
down to low  $\eps$.  Data at lower value of $k$ show a much smother dependence on
$\eps$. The data are compatible with the possibility that for larger systems and for longer
thermalization time a real discontinuity develops. This point deserves to be analysed in
much greater detail. 
On the basis of the date of figure \ref{qeps}, in figure \ref{pha} we 
give a rough estimate the transition line in the plane 
$\Gamma-\eps$ as the line where $q=0.7$.

There are two alternative methods 
which should be give more  accurate estimates of the phase diagram.
\begin{itemize}
\item We start from mixed initial conditions, i.e.  half of the system in the phase with high overlap,
 half of the 
system in the  phase with low overlap  and we study which of the two phases becomes asymptotically stable.
\item We compute the free energy in each of the two phases (apart from a constant) by computing
the internal energy and the overlap along a path that start form a fixed reference point in the 
$\eps -\Gamma$ plane up to the final point.
\end{itemize}

\begin{figure}
\begin{center}
\epsfxsize=350pt
\epsffile{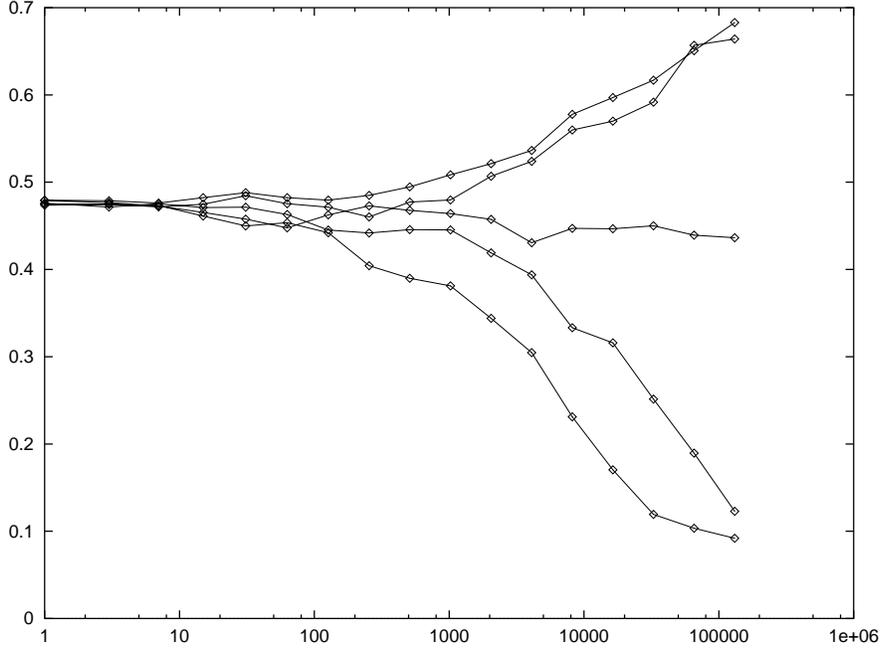} 
\end{center}
   \caption[0]{\protect\label{q135}  
The value of the overlap for different values of $\eps$ (from top to 
bottom $\eps=1.5,1.2,0.9,0.6,0.3$), as a function of Monte Carlo time in a 
logarithmic scale.}
 \end{figure}

In this note we explore only the first procedure. We fist thermalize the system for $2^k$ iterations.
At this point we start
the evolution of a second system which is coupled to the configuration reached by the first one ($y$) 
as in (\ref{Hxy}) with a value of $\eps=4$, taking as  initial condition 
the quenched configuration $y$. The only difference form the previous case is that
we substitute in 
 the Hamiltonian (\ref{Hxy}) $q(x,y)$ by 
\be
q_G(x,y)=1/N \sum_{i,k} w(x_i-y_k) G(x_i^1)
\ee
where $x_i^1$is the first component of the $x_i$ and $G(x)=-1$ for $x<L/2$ and $G(x)=1$ for $x>L/2$. 
In this way we force the overlap of the particles in the first and second half of the box to small and
large values respectively. In this way we have prepared the starting point of the runs done (for others
$2^k$ sweeps) at different 
values of $\eps$ with the usual Hamiltonian (\ref{Hxy}).
In figure \ref{q135} we present the data  for a system of 512 particles for 
$\Gamma=1.35$. We see 
clearly that for high values of $\eps$ the system evolves towards high 
values of $q$ while for small $\eps$, $q$ decreases to low values. The value of 
$\eps$ that separates the two situations
can be estimated to be around $\eps=0.9$ in this figure.

\begin{figure}
\begin{center}
\epsfxsize=350pt
\epsffile{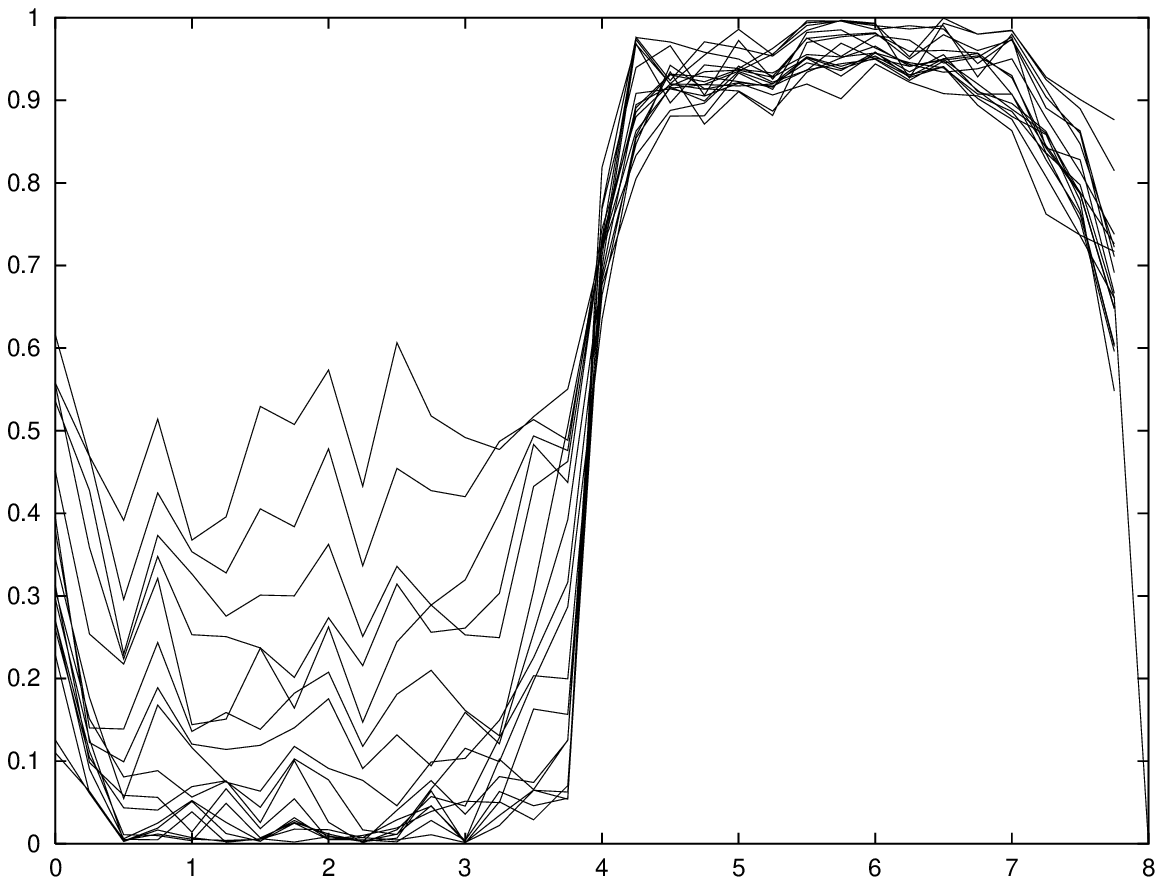} 
\end{center}
   \caption[0]{\protect\label{pro15}  
Time evolution of the space dependent overlap for a system prepared in 
the low overlap phase for $0<x<4$ and in the high overlap phase for $4<x<8$. 
The number of particles is 512, and the box size is 8. The different lines
represent the density profile averaged over $y$ and $z$ at different times
$t_k=2^k$, for $k=0,...,17$. The values of $\Gamma$ and the coupling, 
respectively $\Gamma=1.35$ and $\eps=1.5$ are such that the stable phase 
is the one with high overlap. The curves corresponding to higher times are 
 higher in the low $x$ region. }
 \end{figure}

\begin{figure}
\begin{center}
\epsfxsize=350pt
\epsffile{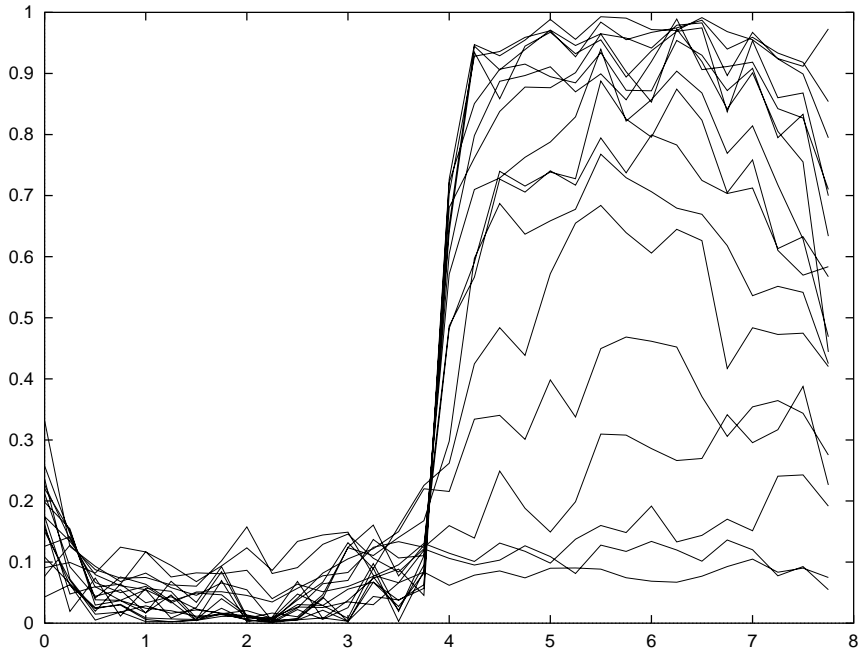} 
\end{center}
   \caption[0]{\protect\label{pro3}  
The same as in figure \ref{pro15} but with a value of the coupling 
($\eps=0.3$) such that the stable phase is the one with low overlap. 
The curves corresponding to higher times are here
lower in the high $x$ region.}
 \end{figure}

The dynamics of equilibration from that initial conditions presents interesting features. 
In figure \ref{pro15} we show the overlap profile at times $t_k=2^k$ for $\eps=1.5$, 
a value such that the high $q$ phase is stable. Figure \ref{pro3} shows the same thing
in a case, $\eps=0.3$, where the low $q$ phase is stable. We notice 
that differently from the usual cases of first order transitions 
where the dominating phase grows at the expenses of the metastable one via a surface
mechanism, here the dominant dynamics seems to occur in the bulk.
The investigation of this kind of dynamics certainly deserves more attention 
than  the one that we have dedicated to it here.

\subsection{Numerical results for the annealed case}

\begin{figure}
\begin{center}
\epsfxsize=350pt
\epsffile{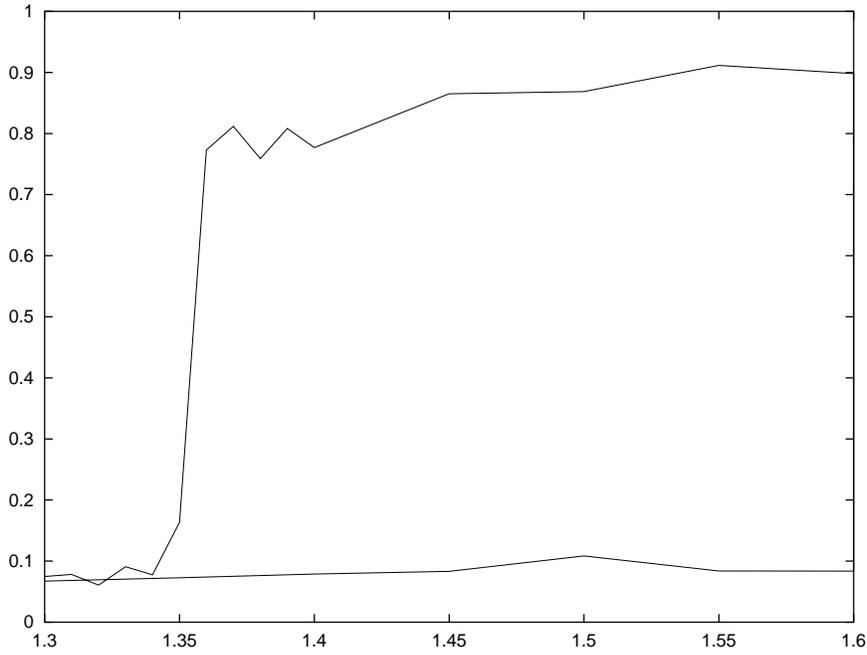} 
\end{center}
   \caption[0]{\protect\label{q15}  
Overlap $q$ for two coupled replicas in the annealed case, during a 
temperature cycle at small coupling $\eps=0.2$ as a function of $\Gamma$. The cooling 
rate is $2^{15}$ step for value of $\Gamma$, the number of particle is 66.  
We start from high temperature $\Gamma=1.3$ we cool
down to $\Gamma=1.6$ (lower curve). At this temperature we set the two configurations
 equal to one of them 
and we heat again (higher curve).
 }
 \end{figure}

\begin{figure}
\begin{center}
\epsfxsize=350pt
\epsffile{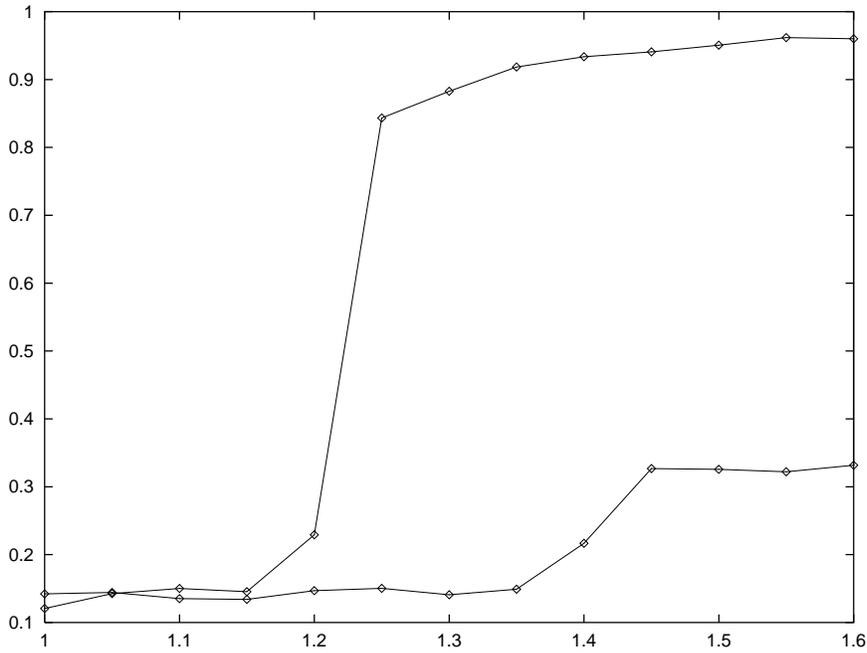} 
\end{center}
   \caption[0]{\protect\label{ann_q}  
Overlap $q$ during a 
temperature cycle as in the previous figure, 
at higher coupling $\eps=0.8$ as a function of $\Gamma$ in the region $\Gamma=1\ -\ 1.6$. 
 }
 \end{figure}

\begin{figure}
\begin{center}
\epsfxsize=350pt
\epsffile{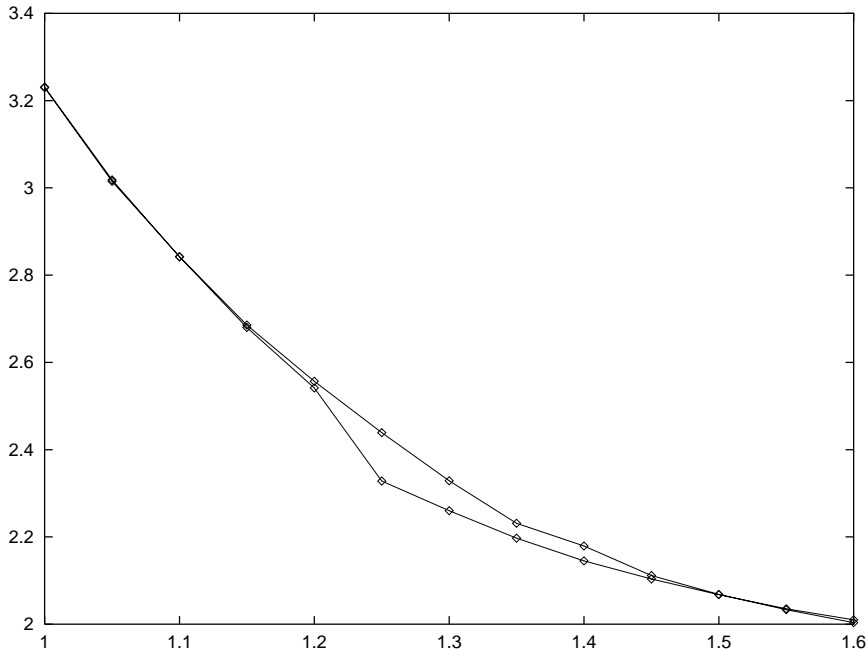} 
\end{center}
   \caption[0]{\protect\label{ann_ene}  
Internal energy  $E$ during the same 
temperature cycle as in the previous figure.
The high curve corresponds to cooling, the lower one to heating.
 }
 \end{figure}

In this case we have two replicas that evolve in parallel with 
Hamiltonian 
\be 
H(x,y)=H(x)+H(y)-N \tilde{\eps} q(x,y). 
\ee
To support our first order transition picture we present
hysteresis data during temperature cycles. The 
procedure consists in first cooling a system in which the two replicas 
start form independent random condition at a low value of $\Gamma$, 
when a maximum value of $\Gamma$ is reached, 
the two configurations are set equal (and equal to one of the two) and the temperature is 
raised again. In figures \ref{q15} and \ref{ann_q} we show data corresponding to 
$\eps=0.2$ and $\eps=0.8$ respectively. We see in both cases that the low $q$ phase 
seems to be metastable for all the probed values of the temperature. On heating the system
passes from the high $q$ phase to the low $q$ phase with a sharp discontinuity. 
In figure \ref{ann_ene} we present data for the internal 
energy for the same cycle of figure \ref{ann_q}. 
The hysteretic behavior found there is an important indication of a first order 
phase transition. Notice that the high $q$ phase, where the entropy is lower, 
has a lower 
internal energy and is present also in the 
 liquid phase for $\Gamma<\Gamma_c$. Indeed the difference in energy 
 starts to be present just around $\Gamma_c$. This behavior is in agreement 
 with the theory. Low energy metastable states exist also for $T>T_c$, 
 but they have zero Boltzmann weight in that region. Coupling two 
 replica together enhance their probability so that their existence can 
 be observed \cite{KPV,BBM,bfp,I}. 

\section{Conclusions}

In this paper we have given theoretical arguments and numerical support 
in favor of a description of the ideal glassy transition as the limiting 
point of a first order transition line in the plane of the temperature and
the coupling among real replicas.

 Similar conclusions are obtained 
in the quenched construction and the annealed one. The first construction
studies the implications of a Boltzmann-Gibbs distribution limited to
the configurations which have a fixed overlap with a quenched configuration. 
The second construction studies 
 a Boltzmann-Gibbs distribution of two systems on the same foot 
and with  fixed overlap. 

In both formalisms the glassy state is described as a state where 
there are two phases. A ``confined phase'' with high value of the overlap, and 
a ``deconfined phase'' where there  is minimal correlation. We have shown that
the glassy transition, which is Ehrenfest second-order for zero coupling, 
becomes first order as soon as a non zero coupling is introduced.
A detailed computation in the example of the spherical p-spin model has shown that 
 that there  is a 
first order transition line in the $\eps-T$ plane terminating in a critical 
point.
This result should be robust beyond mean field. 
Indeed, we have discussed how
the numerical simulations for binary soft-sphere mixtures 
at non zero values of the coupling $\eps$ support the theoretical picture. 
A much greater numerical effort would however be needed to locate with precision 
the transition line and to study the interesting issue of the nature 
of the glassy critical point. 
A key point in  our analysis, in going from infinite range to short range models, 
is the possibility to use the Maxwell construction to estimate the 
topology of the transition line. In ordinary systems 
the validity of the Maxwell construction is intimately related 
to the phenomenon of phase coexistence.  
In the case of glasses we do not know what a supposed  coexistence 
of the confined and the deconfined phase would mean.
Our numerical simulations, which seem to  confirm the first order transition 
picture, indicate an overlap dynamics different from the usual domain growth 
in spinodal decomposition. This question, and the one  of the decay of the 
high $q$ phase in the metastable regime are deeply related to the problem 
of restoring of ergodicity when the barriers are finite. This is at the heart 
of the glassy physics and further numerical and theoretical 
effort is certainly needed.

\section*{Acknowledgments} S.F. thanks the ``Dipartimento di Fisica dell' Universit\`a
di Roma La Sapienza'' for kind hospitality during the elaboration of this work.

\end{document}